\documentclass[aps,physrev,reprint,superscriptaddress,longbibliography,floatfix]{revtex4-2}
\usepackage{amsmath}
\usepackage{amssymb}
\usepackage{xcolor}
\usepackage[english]{babel}
\usepackage{times}
\usepackage{hyperref}
\usepackage{soul}
\usepackage[table]{xcolor}
\usepackage{multirow}
\usepackage{xcolor}
\definecolor{orcidgreen}{RGB}{166,206,57}

\usepackage{graphicx}
\usepackage{fontawesome5}

\usepackage{hyperref} 
\hypersetup{
  colorlinks=true,
  linkcolor=blue,
  filecolor=black,
  urlcolor=blue,
  citecolor=blue
}

\newcommand{\orcidicon}[1]{%
  \href{https://orcid.org/#1}{\texorpdfstring{\textcolor{orcidgreen}{\faOrcid}}{\faOrcid}}%
}
\begin{document}

\title{{\color{black}Long-range mid-infrared energy transfer mediated by hyperbolic phonon polaritons}}

\author{Gonzalo \'Alvarez-P\'erez\ \orcidicon{0000-0002-4633-1898}}
\email{gonzalo.alvarezperez@iit.it}
\affiliation{Istituto Italiano di Tecnologia, Center for Biomolecular Nanotechnologies, Via Barsanti 14, 73010 Arnesano, Italy}
\author{Simone De Liberato\ \orcidicon{0000-0002-4851-2633}}
\affiliation{Istituto di Fotonica e Nanotecnologie, Consiglio Nazionale delle Ricerche (CNR), Piazza Leonardo da Vinci 32, Milano, 20133, Italy}
\affiliation{School of Physics and Astronomy, University of Southampton, Southampton SO17 1BJ, United Kingdom}
\author{Huatian Hu\ \orcidicon{0000-0001-8284-9494}} 
\email{huatian.hu@iit.it}
\affiliation{Istituto Italiano di Tecnologia, Center for Biomolecular Nanotechnologies, Via Barsanti 14, 73010 Arnesano, Italy}
\date{\today}

\begin{abstract}
We provide a framework to theoretically describe long-range energy transfer in single and twisted two-dimensional hyperbolic slabs. We demonstrate that phonon polaritons (PhPs)—quantum superpositions of photons and lattice vibrations in polar dielectrics—can mediate and enhance room-temperature energy transfer at ranges far exceeding those of conventional mid-infrared (MIR) platforms, and with extreme directionality. {\color{black}This is because the dipole-dipole interaction potential energy diverges along the asymptotes of the real-space hyperbolic opening angle. Our findings} allow us to extend classical and quantum interactions between dipoles, typically strictly confined to the near-field, beyond several free-space MIR wavelengths. We use $\alpha$-MoO\textsubscript{3} as a representative material, but this mechanism is not limited to the MIR---it is general to anisotropic media across the whole electromagnetic spectrum.
\end{abstract}

\maketitle

\textit{Introduction}---Energy transfer between localized excitations—such as atoms, molecules, or spins—is a universal process governing interactions across the electromagnetic spectrum. It can occur radiatively through photon emission and reabsorption or nonradiatively via direct coupling. Resonant energy transfer stands out as a fundamental pathway in which a donor transfers energy nonradiatively to an acceptor through so-called dipole-dipole interactions (DDIs), mediated by virtual photon exchange. DDIs underpin a broad range of phenomena: from Casimir and van der Waals (vdW) forces \cite{rodriguez2011casimir,Cortes17super}, superradiance and entanglement in atomic ensembles \cite{vanLoo13photon,Goban15superradiamce,Hubner98superradiant}, and Förster energy transfer (FRET) between molecules or quantum dots \cite{Andrew00forster,Shahmoon13nonradiative,Kang25long}, to cooperative shifts in superconducting qubits and magnonic systems \cite{Kweon94resonance,Douglas15quantum,Cornelissen15long,Luthi25long,Liu18long}.

\begin{figure}[ht]
\includegraphics[width=0.5\textwidth]{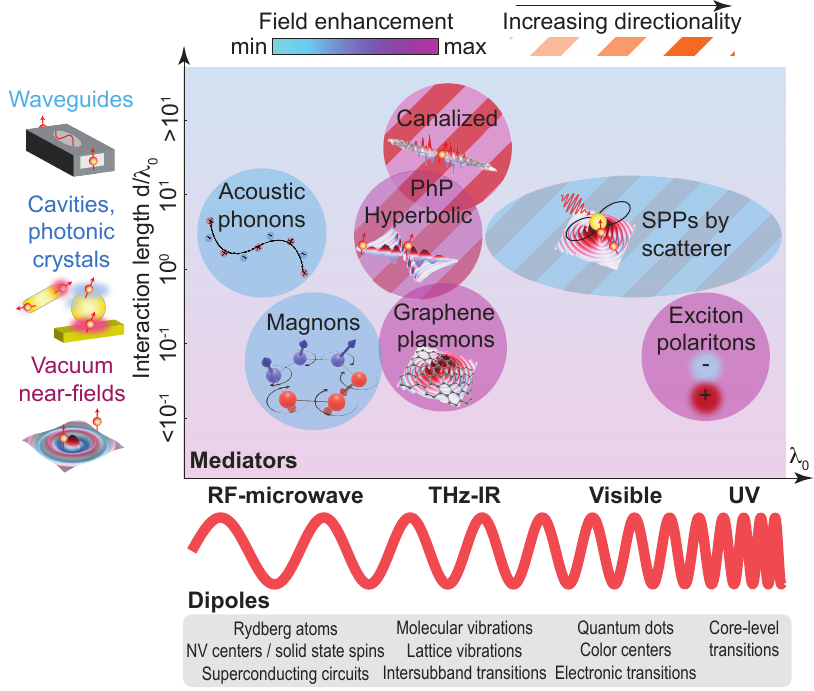}
\caption{\label{fig1}
\textbf{Landscape of platforms for long-range DDIs across the electromagnetic spectrum.} DDI mediators are compared by their normalized interaction length $d/\lambda_0$. Platforms are arranged by operating frequency. Shading denotes relative field enhancement and orange stripes indicate directionality.
}\end{figure}

However, their strength decays rapidly with distance—scaling as $1/r^3$ in the near field and $1/r$ in the far field \cite{Newman18observation}—restricting coherent coupling to subwavelength separations. Extending their range requires a mediator capable of carrying electromagnetic energy beyond the donor’s near field. Two main strategies are typically employed for this: tuning the emitters’ intrinsic properties (e.g., Rydberg atoms, superconducting qubits) \cite{milonni94quantum,Poddubny13hyperbolic,John91quantum}, or engineering the mediating platforms to tailor propagation and coupling, using cavities, waveguides, or metamaterials \cite{boddeti2024reducing,Newman18observation,Milton13searchlight,Tanaka10multifold,Sokhoyan13quantum,Vesseur13experimental}. These mediating platforms, summarized in Fig. \ref{fig1}, span all spectral regimes and show the fundamental trade-offs between field enhancement, interaction length, and directionality. Vacuum near-fields dominate at subwavelength scales, providing strong but highly localized coupling. Cavities and photonic crystals enhance coherence through mode confinement but remain spectrally narrow, while waveguides enable long-range and directional energy transfer with moderate field enhancement. Importantly, collective excitations can also act as frequency-specific mediators: magnons in the RF–microwave range, acoustic phonons in the THz domain, graphene plasmons and surface plasmon polaritons (SPPs) in the mid-IR to visible, and exciton polaritons in the visible to UV. In particular, localized surface plasmons (LSPs) can strongly enhance near fields at visible frequencies but are limited to interaction lengths of $d/\lambda_0 \sim 0.1 - 1$ \cite{hamza2021forster,hamza2023long,hsu2017plasmon}, while SPPs and graphene plasmons can extend the interaction range but with a lower confinement with respect to LSPs and
less directional control than waveguides \cite{Hu22nanoparticle,Li20duplicating,fang2009remote,bouchet2016long,Huidobro12,Kang25long}. As such, no existing platform simultaneously achieves strong confinement, long-range {\color{black}interaction}, and high directionality.

These trade-offs become particularly critical in the MIR (3–20 $\mu$m), where light couples to molecular vibrations and infrared-active phonons rather than to electronic or magnonic excitations. In this regime, energy exchange occurs in the form of vibrational energy transfer (VET), which is governed by vibrational DDIs \cite{Xiang20intermolecular,Yu22intermolecular,Newman18observation}. Energy can be coherently exchanged between vibrational excitations despite {\color{black}they} are thermally populated and short-lived at room temperature. In fact, controlling such process is important for the nanoscale control of infrared energy flow. Additionally, while efficient quantum emitters are scarce in this range, coupling molecular vibrations, intersubband transitions, or optical phonons via long-range DDIs offers a promising route to manipulate vibrational interactions—much as plasmonics enabled light–matter control before the advent of single-photon emitters. Yet, like their electronic and magnonic counterparts, vibrational DDIs decay rapidly with distance, preventing coherent coupling between spatially separated dipoles. Existing mediating strategies fail to overcome these limitations: metal-based plasmons suffer from high losses and poor confinement, graphene plasmons have stronger confinement but still have rather high ohmic losses, dielectric resonators offer modest enhancement without directionality, and microcavities are constrained by narrow bandwidths and fabrication complexity \cite{Hu22nanoparticle,Kang25long}. Consequently, enhancing vibrational DDIs beyond the near field in the MIR remains an open challenge.

In this context, phonon polaritons (PhPs) in polar dielectrics offer a promising solution. These hybrid excitations---arising from the strong coupling between infrared photons and optical phonons---combine deep subwavelength confinement, low {\color{black}optical} losses, and strong field enhancement in the MIR \cite{Caldwell15low}. Crucially, PhPs spectrally overlap with the vibrational modes of numerous molecular species, enabling vibrational strong coupling. This phenomenon has been experimentally demonstrated in several studies, both in the far field \cite{Autore2018_boron} and the near field \cite{Bylinkin21_real,TresguerresMata25_directional}, allowing remote \cite{Dolado22_remote} and on-chip \cite{Bylinkin24_onchip} detection of molecular vibrations. Moreover, anisotropic and hyperbolic PhPs in low-symmetry polar dielectrics enable extreme confinement and directional propagation \cite{Galiffi24extreme,kowalski25_ultraconfined}, allowing dipoles to interact far beyond the natural $1/r^3$ range. It has been recently demonstrated that hyperbolic PhPs in graphene/hBN heterostructures can transfer a significant fraction of the injected electronic and thermal power out of plane via mid-infrared radiation \cite{AbouHamdan25}. This highlights the potential of hyperbolic PhPs for strong light-matter interactions with low losses, but their capabilities as long-range energy mediators remain unexplored.

Here, we theoretically demonstrate that hyperbolic and canalized PhPs can in fact serve as exceptional mediators of long-range DDIs in the MIR regime, as they combine strong field confinement, {\color{black}low optical losses}, pronounced directionality and large local density of optical states (LDOS). {\color{black}These} properties enable vibrational dipoles to couple over tens of micrometres—distances exceeding $5\lambda_0$ and orders of magnitude larger than the PhP wavelength in the MIR. Remarkably, the DDI strength is enhanced by factors greater than $10^3$ compared with gold- or SiC-based platforms, and even surpasses that achieved with graphene plasmons. Moreover, we show that twisting $\alpha$-MoO\textsubscript{3} layers allows to tune the dipole–PhP coupling and thus long-range DDIs in the MIR. Our formalism provides a general framework applicable to strongly anisotropic materials throughout the electromagnetic spectrum.

\begin{figure}[!ht]
\includegraphics[width=0.5\textwidth]{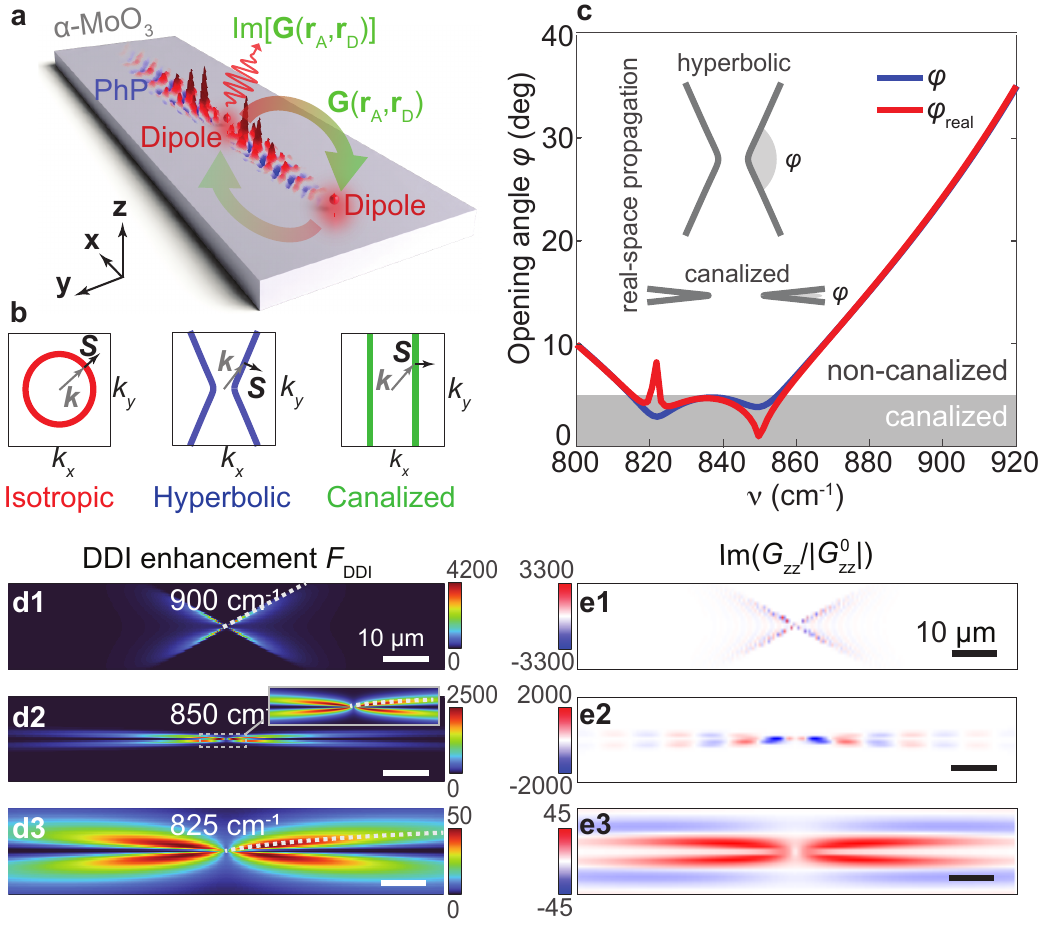}
\caption{\label{fig2}
\textbf{Long-range MIR DDIs mediated by hyperbolic PhPs in a single $\alpha$-MoO\textsubscript{3} slab.}
\textbf{(a)} Two dipoles on a 200-nm-thick $\alpha$-MoO\textsubscript{3} slab, interacting via PhPs described by the DGF $\tilde{\mathbf{G}}(\mathbf{r}_A,\mathbf{r}_D)$.
\textbf{(b)} IFCs for isotropic (left), hyperbolic (middle), and canalized (right) PhPs; grey and black arrows indicate $\mathbf{k}_p$ and the propagation direction $\mathbf{S}$.
\textbf{(c)} Opening angle $\varphi$ of hyperbolic PhP propagation from 800 to 920~cm\textsuperscript{-1}, calculated using complex (blue) and real-part (red) permittivities; canalization occurs around 820–850~cm\textsuperscript{-1}.
\textbf{(d1–d3)} Normalized DDI enhancement $F_{\rm DDI}$ at 900, 850, and 825~cm\textsuperscript{-1} for a 200-nm $\alpha$-MoO\textsubscript{3} slab.
\textbf{(e1–e3)} Normalized imaginary part $\mathrm{Im}[G_{zz}/|G_{zz}^0|]$ at the same frequencies, showing phase and LDOS patterns.
}\end{figure}

\textit{Long-range MIR DDIs mediated by hyperbolic PhPs in a single slab}---Some vdW materials display strong structural anisotropy \cite{Lajaunie13strong}. {\color{black}This typically leads to strongly direction-dependent optical responses. As a result, light propagation in such crystals is described by a frequency-dependent dielectric tensor $\hat{\varepsilon}(\nu) = \mathrm{diag}[\varepsilon_x(\nu), \varepsilon_y(\nu), \varepsilon_z(\nu)]$} \cite{Ma18inplane,Zheng19mid,AlvarezPerez20infrared}. Importantly, the signs of the in-plane permittivity components $\varepsilon_{x,y}(\nu)$ mainly determine the propagation along the material's surface. This can be visualized by plotting the isofrequency curve (IFC), i.e., the polariton dispersion $\nu(\mathbf{k})$ with in-plane wavevector $\mathbf{k}=[k_x,k_y]$. When $\varepsilon_x=\varepsilon_y$, propagation is isotropic and the IFC is circular (Fig. \ref{fig2}b, left), with the Poynting vector $\mathbf{S}$—perpendicular to the IFC \cite{Voronin24fundamentals,Voronin2025}—indicating uniform in-plane propagation. When $\varepsilon_x$ and $\varepsilon_y$ have opposite signs (within the reststrahlen bands bounded by TO and LO phonons \cite{adachi99optical}), the IFC becomes hyperbolic, restricting propagation to specific directions (Fig. \ref{fig2}b, middle). Such in-plane hyperbolic behavior occurs naturally in several vdW \cite{TaboadaGutierrez20broad,Ruta23hyperbolic,Venturi24visible,ruta25good} and bulk \cite{Ma21ghost,Passler22hyperbolic} crystals, as well as in hyperbolic metamaterials \cite{cortes2018fundamental,Cortes17super}. Wavevectors near the hyperbola’s asymptotes correspond to propagation along a single direction, yielding a high LDOS. Interestingly, stacking in-plane hyperbolic vdW layers at different twist angles allows to control the polariton hybridization and propagation, from the MIR \cite{Hu20topological,Duan20twisted,Chen20configurable,Zheng20phonon} down to the terahertz \cite{Obst2023_terahertz}. At a specific critical twist angle, the so-called \textit{magic} angle, the IFC flattens, leading to canalization, defined by all the allowed modes propagating collinearly without diffraction (Fig. \ref{fig2}b, right). This regime has been demonstrated in twisted bilayers \cite{Hu20topological,Duan20twisted,Zheng20phonon,Chen20configurable} and trilayers \cite{Duan23multiple} of $\alpha$-MoO\textsubscript{3}, a {\color{black}paradigmatic} MIR in-plane hyperbolic material. 

Figure \ref{fig2}a illustrates two dipoles, a donor and an acceptor supporting MIR vibrational modes and coupled via PhPs on a $\alpha$-MoO\textsubscript{3} slab. The dipoles are located at $\mathbf{r}_j$ ($j = D, A$) and have transition frequencies $\omega_j$ and transition dipole moments $\hat{\boldsymbol{\mu}}_j$. The donor’s excitation can be non-radiatively transferred to the acceptor via DDIs \cite{Novotny12principles}, without photon emission. The resonant DDI potential energy governing the energy transfer between the donor and the acceptor can be derived within quantum electrodynamics (QED). The detailed derivation of the DDI potential energy enhancement is performed in Appendix I. We set the slab thickness $d=200$ nm, the donor and acceptor height over the slab to $H=20$ nm, and assume the half-spaces over and under the slab to be filled by vacuum ($\varepsilon = 1$). We focus on the spectral range 800–920 cm$^{-1}$ (12.5–10.87 $\mu$m), where $\varepsilon_x < 0$ and $\varepsilon_y > 0$, leading to in-plane hyperbolic dispersion \cite{AlvarezPerez20infrared} (see Supplementary Section I). The propagation angle $\varphi = \arctan(i \sqrt{\varepsilon_y/\varepsilon_x})$ (or $\varphi_{\rm real} = \arctan[i \sqrt{\mathrm{Re}(\varepsilon_y)/\mathrm{Re}(\varepsilon_x)}]$) \cite{AlvarezPerez19analytical, Dai15Subdiffractional, Li15Hyperbolic} describes the opening of the hyperbolic rays (Fig.~\ref{fig2}b). Between 800 and 920 cm$^{-1}$, canalization occurs at lower frequencies, while $\varphi$ increases roughly linearly above 860 cm$^{-1}$. Two minima in $\varphi$ appear at 825 and 850 cm$^{-1}$—the latter corresponding to standard canalization, and the former to \emph{loss-induced canalization} \cite{TresguerresMata24Observation,TeranGarcia25real}.

Figures~\ref{fig2}d1–d3 display $F_{\rm DDI}$, the normalized DDI enhancement, for various frequencies. The first dipole is fixed at the center, and the second varies along the $x$–$y$ plane. $F_\text{DDI}\leq 1$ indicates weaker coupling than in vacuum.
At $\nu_0=900$ cm$^{-1}$ (Fig.~\ref{fig2}d1), $F_{\rm DDI}$ peaks along the hyperbolic asymptotes ($F_{\rm DDI}\approx4200$).  {\color{black}To illustrate the mechanism for this enhancement, let us rewrite the last term in the denominator of Eq.~\eqref{eq:Gzz} as $\varepsilon_x \Delta y^2 + \varepsilon_y \Delta x^2 = r^2 \cos^2\varphi (\varepsilon_x \tan^2\varphi + \varepsilon_y )$. As the in-plane angle $\varphi$ approaches the opening angle $\varphi \rightarrow \varphi_c = \arctan(i \sqrt{\varepsilon_y/\varepsilon_x})$, we have
$r^2 \cos^2\varphi (\varepsilon_x \tan^2\varphi + \varepsilon_y ) \rightarrow r^2 \cos^2\varphi \left[ \varepsilon_x (-\varepsilon_y/\varepsilon_x) + \varepsilon_y \right] = 0$. As such, as $\varphi \rightarrow \varphi_c$, $G_{zz}(\mathbf{r}_A - \mathbf{r}_D) \rightarrow \infty$ and therefore $V_{dd}(\mathbf{r}_A - \mathbf{r}_D) \rightarrow \infty$. Consequently, the in-plane hyperbolicity of the system leads to a divergence of the DDI potential energy along the real-space hyperbolic opening angle $\varphi_c$, even if losses are higher along this direction (see Supplementary Section II).} The imaginary part $\mathrm{Im}(G_{zz}/G_{zz}^0)$ (Fig.~\ref{fig2}e1) reflects the PhP phase and LDOS enhancement, peaking along the asymptotes. At $\nu_0=850$ cm$^{-1}$ (Fig.~\ref{fig2}d2), $F_{\rm DDI}$ reaches $\sim2500$, lower but with longer {\color{black}interaction} distances due to reduced $\mathrm{Im}(\varepsilon)$. Here, $\mathrm{Re}(\varepsilon_x)\approx -22$ and $\mathrm{Re}(\varepsilon_y)\approx 0$, corresponding to an ENZ point that induces canalized propagation along $x$ \cite{Bai25direct}. Although the enhancement is weaker, the strong directionality enables controlled nanoscale energy transport (Fig.~\ref{fig2}e2). At $\nu_0=825$ cm$^{-1}$ (Fig.~\ref{fig2}d3), $F_{\rm DDI}\approx50$. The permittivities are $\varepsilon_x=127+109i$ and $\varepsilon_y=-0.6+0.12i$, producing highly directional, loss-induced canalization along $x$ \cite{TresguerresMata24Observation,TeranGarcia25real}. The phase remains nearly constant, in what has been named \emph{ray canalization} (Fig.~\ref{fig2}e3), enabling the longest-range coupling with minimal decay, though at the cost of reduced enhancement.

\begin{figure}[!ht]
\includegraphics[width=0.3\textwidth]{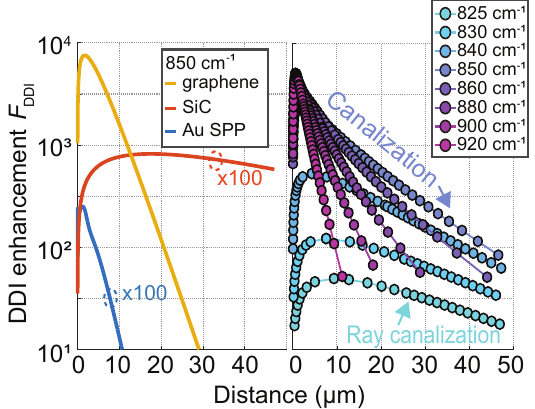}
\caption{\label{fig2f_revised}
{\color{black}\textbf{Benchmarking long-range MIR DDIs mediated by hyperbolic PhPs in a single $\alpha$-MoO\textsubscript{3} slab against existing MIR platforms.}
\textbf{Left:} $F_{\rm DDI}$ for SPPs on Au, SPhPs in SiC, and graphene plasmons. \textbf{Right:} $F_{\rm DDI}$ in a 200-nm $\alpha$-MoO\textsubscript{3} slab along the direction of maximum enhancement in the 825–920~cm\textsuperscript{-1} range.}
}\end{figure}

This trade-off between strength and range is summarized in Fig.~\ref{fig2f_revised}, which further shows that hyperbolic PhPs in $\alpha$-MoO\textsubscript{3} (825–920 cm$^{-1}$) provide a promising platform to mediate strong and ultra-long-range DDIs in the MIR, far surpassing conventional MIR platforms such as bare gold and SiC 200-nm-thick slabs, or a single graphene monolayer ($E_F=0.3$ eV). Hyperbolic PhP-mediated DDIs are strong and can extend up to several PhP wavelengths ($\approx\lambda_0=$10 $\mu$m). At lower frequencies, the enhancement weakens but persists over $\sim$50 $\mu$m ($\approx 5\lambda_0$), where $F_{\rm DDI}>10$. The optimal balance occurs at 850 cm$^{-1}$, providing a 1000-fold enhancement and 5$\lambda_0$ range. Analytical calculations of the {\color{black}interaction} length confirm this trend and show that the dependence on the dipole height $H$ is weak (see Supplementary Section {\color{black}III}). For even longer-range interactions ($>$50 $\mu$m), ray canalization (825–840 cm$^{-1}$) becomes advantageous despite its lower short-range enhancement.

Interestingly, Eq.~\eqref{dyadic_GF_3} can be expressed as \(F_\text{DDI}^0 \exp(-R/L_p)\) (see Supplementary Section {\color{black}IV}), where \(F_\text{DDI}^0\) denotes the local enhancement of the DDI potential energy at the dipole position, with its real part corresponding to the Lamb shift and its imaginary part to the Purcell factor \cite{Novotny12principles}. The quantity $F_\text{DDI}^0 \exp(-R/L_p)$ thus completely encapsulates the aforementioned trade-off between long-range coupling and DDI potential enhancement. By integrating the \(F_\text{DDI}\) curves shown in Fig.~\ref{fig2f_revised}, we obtain single data points that capture both, revealing that the optimal configuration occurs at 850~cm\(^{-1}\) {\color{black}(see Supplementary Figure 4).}

\begin{figure}[ht]
\includegraphics[width=0.45\textwidth]{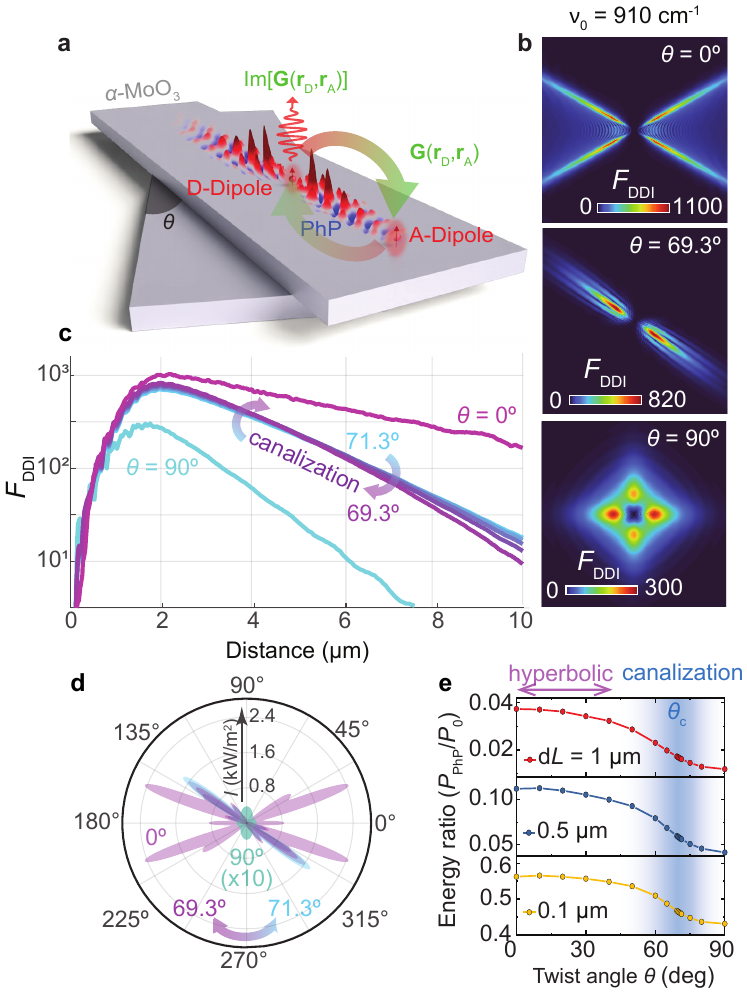}
\caption{\label{fig3}
\textbf{Long-range MIR DDIs mediated by canalized PhPs in twisted bilayer $\alpha$-MoO\textsubscript{3}.}
\textbf{(a)} Two dipoles on twisted $\alpha$-MoO\textsubscript{3} slabs at angle $\theta$, interacting via PhPs described by the DGF $\tilde{\mathbf{G}}(\mathbf{r}_A,\mathbf{r}_D)$.
\textbf{(b)} Normalized DDI enhancement $F_{\rm DDI}$ at 910~cm\textsuperscript{-1} for twist angles $\theta=0^\circ$, $69.3^\circ$, and $90^\circ$.
\textbf{(c)} $F_{\rm DDI}$ along the direction of maximum coupling (canalization or asymptotes) for the same system.
\textbf{(d)} Real-space PhP intensity patterns as a function of twist angle, obtained analytically from the DGF.
\textbf{(e)} Ratio of PhP power $P_{\rm PhP}$ to total dipole decay $P_0$ versus twist angle for several distances from the donor.
} \end{figure}

\textit{Long-range MIR DDIs mediated by canalized PhPs in twisted biayers}---Given the exceptionally strong long-range DDIs mediated by hyperbolic PhPs in single slabs, we now study DDIs in two twisted  slabs rotated by an angle $\theta$ (Fig.~\ref{fig3}a), using $\alpha$-MoO$_3$ as a representative example. To accurately compute the DDI enhancement factor $F_{\rm DDI}$, we perform full-wave 3D finite-element method (FEM) simulations (\textsc{Comsol Multiphysics}), since 2D approximations fail to capture the out-of-plane coupling relevant in twisted systems~\cite{Duan20twisted,Duan23multiple}. Each slab is 100~nm thick, matching the 200~nm single-layer case in Fig.~\ref{fig2}. We evaluate $F_{\rm DDI}$ at $\nu_0=910$~cm$^{-1}$ ($\approx$11~$\mu$m) for three twist angles: $\theta=0^\circ$, $69.3^\circ$, and $90^\circ$. 

At $\theta=0^\circ$ (Fig.~\ref{fig3}b, left), the enhancement pattern resembles that of the 200~nm single slab, with additional fine fringes near the origin from high-order modes~\cite{AlvarezPerez19analytical}. Quantitative differences arise because the previous case used a 2D model, whereas here we carry out a full 3D simulation. The {\color{black}interaction} lengths are $L_p=3.5~\mu\mathrm{m}$ along the hyperbola’s axis and $L_p=2.5~\mu\mathrm{m}$ along its asymptotes, as derived from the analytical dispersion \cite{Duan23multiple} (see Supplementary Section {\color{black}V}). At $\theta=69.3^\circ$ (Fig.~\ref{fig3}b, middle), PhPs become canalized, propagating more directionally but with lower intensity and shorter {\color{black}interaction} length ($L_p=1.96~\mu\mathrm{m}$). This trend intensifies at $\theta=90^\circ$ (Fig.~\ref{fig3}b, right), where propagation occurs in all directions with reduced range ($L_p=1~\mu\mathrm{m}$) and slightly asymmetric modes due to the broken symmetry between the layers. Figure~\ref{fig3}c shows $F_{\rm DDI}$ versus distance for different $\theta$. The hyperbolic case ($\theta=0^\circ$) yields the largest enhancement ($F_{\rm DDI}\approx1100$) and longest range, while at $\theta=90^\circ$ the enhancement drops to $\approx300$. Near the so-called \textit{magic angle} ($\theta\approx69$--$71^\circ$), the enhancement is slightly lower than in the hyperbolic case at all distances, despite the increased directionality. This behavior is consistent with Fig.~\ref{fig3}b, where canalization produces the most directional but not the strongest DDI patterns. The analytical polar plots of the PhP intensity maps at $5L_p$ (Fig.~\ref{fig3}d) further confirm this (see Supplementary Section {\color{black}VI} for the derivation), showing the balance between extremely directional {\color{black}interaction} and long-range DDIs as a function of the twist angle, which provides a tuning knob for controlling MIR energy transfer.

To understand why canalized PhPs do not yield the strongest long-range DDIs, we analyze the dipole’s energy distribution among three decay channels: (1) guided PhPs, (2) negligible free-space radiation, and (3) localized absorption (quenching). Since rigorous near-to-far-field transformations are challenging in anisotropic media \cite{yang2016near}, we estimate the PhP power $P_{\rm PhP}$ from the Poynting vector’s surface integral over a cylindrical shell of radius $dL$ and height 800~nm, excluding top and bottom surfaces to isolate in-plane energy flow. The ratio $P_{\rm PhP}/P_0$, where $P_0$ is the total dipole decay power, measures the in-coupling efficiency of PhPs (Fig.~\ref{fig3}e). For all distances, $P_{\rm PhP}/P_0$ decreases as $\theta$ increases, with a sharp drop near the canalization (magic) angle $\theta_c$, explaining the trade-off between directionality and coupling strength. As distance grows, the ratio further declines due to {\color{black}optical} losses. Since LDOS reflects available photonic states rather than coupling efficiency, the reduced $P_{\rm PhP}/P_0$ near $\theta_c$ arises from the lower LDOS, not weaker coupling.

In conclusion, canalization enhances the propagation directionality but inherently reduces the LDOS and in-coupling efficiency. In hyperbolic slabs ($\varepsilon_x\varepsilon_y<0$), the interaction potential $|V_{\text{dd}}|$ diverges along the IFC asymptotes, producing super-Coulombic enhancement. However, canalized PhPs lack such divergence, yielding smaller $|V_{\text{dd}}|$. This provides, for the first time, a quantitative explanation of how free-space radiation couples into PhP canalization---an insight crucial for future directional nanophotonics applications~\cite{deabajo2025roadmap}.

\textit{Summary and outlook}---We have provided a theoretical framework to describe long-range DDIs in strongly anisotropic media. Using single and twisted $\alpha$-MoO\textsubscript{3} slabs as representative strongly anisotropic materials, we have shown that low-loss hyperbolic PhPs can mediate exceptionally strong and long-range DDIs in the MIR, with extreme directionality, surpassing traditional MIR platforms—SPPs on gold and SPhPs in SiC—by more than three orders of magnitude, as well as those achieved by graphene plasmons. The coupling range can extend to beyond 50 $\mu$m (about five free-space wavelengths). We have also shown that canalization enhances directionality by confining propagation to a single in-plane direction, but this restriction inherently limits the in-coupling efficiency due to the reduced density of accessible states. Our study thus establishes a new paradigm for strong, long-range MIR DDIs and provides a quantitative framework for understanding the coupling of free-space radiation into canalized PhPs, highlighting the interplay between confinement, long-range interactions and directionality. These insights open avenues for directional vibrational energy transport, thermal management, nanoscale sensing, and MIR quantum information technologies. {\color{black}Our framework is general and provides a fundamental basis to achieve strong long-range DDIs and energy transfer across the electromagnetic spectrum, enabled by the expanding class of in-plane hyperbolic materials and their twisted structures, supporting hyperbolic plasmon polaritons in the visible \cite{Venturi24visible,ruta25good}, hyperbolic plasmon, exciton, and phonon polaritons in the near- and mid-infrared \cite{Galiffi24extreme,Wang2020_vanderwaals,Ruta23hyperbolic}, and hyperbolic phonon polaritons in the far-infrared \cite{deOliveira2021_nanoscale}.}

\textit{Acknowledgments}---We thank Kirill Voronin, Alexey Nikitin, Thomas G. Folland, Alexander Paarmann and Pablo Alonso González for helpful discussions. {\color{black}G.Á.-P. acknowledges support from the European Union (Marie Skłodowska-Curie Actions, grant agreement No. 101209198).} S.D.L. acknowledges financial support under the National Recovery and Resilience Plan (NRRP), Mission 4, Component 2, Investment 1.1, Call for tender No. 1409 published on 14/09/2022 by the Italian Ministry of University and Research (MUR), funded by the European Union – NextGenerationEU – Project Title MINAS - CUP B53D23028420001 - Grant Assignment Decree No. 1380 adopted on 01/09/2023 by the Italian Ministry of University and Research (MUR). {\color{black}Views and opinions expressed are those of the authors only and do not necessarily reflect those of the European Union. Neither the European Union nor the granting authority can be held responsible for them.}

\section*{Appendix I: Calculation of the DDI potential energy enhancement}

Here we derive the DDI potential energy enhancement between two dipoles located on a slab of a biaxial material. Specifically, the interaction between the two dipoles described through the Hamiltonian
\[
\hat{\mathcal{H}}_{\mathrm{int}} = - \sum_j \int_0^\infty d\omega \left[ \hat{\boldsymbol{\mu}}_j \cdot \hat{\mathbf{E}}(\mathbf{r}_j, \omega) + \mathrm{h.c.} \right],
\]
where h.c. denotes the Hermitian conjugate. The quantized electric field in an absorbing, dispersive medium is then expressed in terms of bosonic operators $\hat{\mathbf{a}}(\mathbf{r}, \omega)$ and $\hat{\mathbf{a}}^\dagger(\mathbf{r}, \omega)$, which create and annihilate polaritonic excitations:
\[
\hat{\mathbf{E}}(\mathbf{r}_A, \omega) = i \sqrt{\frac{\hbar}{\pi \varepsilon_0}} \frac{\omega^2}{c^2} \int d^3\mathbf{r}_D \sqrt{\hat{\varepsilon}} \, \tilde{\mathbf{G}}(\mathbf{r}_A, \mathbf{r}_D; \omega) \cdot \hat{\mathbf{a}}(\mathbf{r}_D, \omega),
\]
where $\varepsilon_0$ is the vacuum permittivity, $c$ is the speed of light, and $\tilde{\mathbf{G}}(\mathbf{r}_A, \mathbf{r}_D; \omega)$ is the dyadic Green’s function (DGF), a symmetric $3\times3$ tensor that governs field propagation \cite{Novotny12principles}.

Each dipole ($D$ or $A$) is treated as a two-level system with ground and excited states $|g_{A,D}\rangle$ and $|e_{A,D}\rangle$. For a donor at $\mathbf{r}_D$, the interaction Hamiltonian is $\hat{\mathcal{H}}_D = -\hat{\boldsymbol{\mu}}_D \cdot \hat{\mathbf{E}}(\mathbf{r}_D)$, and its de-excitation is described by the quantum current density operator  
$\hat{\mathbf{j}}(\mathbf{r}_A; \omega) = -i\omega \, \delta(\mathbf{r}_A, \mathbf{r}_D) \, \boldsymbol{\mu}_D \, \hat{\sigma}_D$,  
where $\boldsymbol{\mu}_D = \langle g_D | \hat{\boldsymbol{\mu}}_D | e_D \rangle$ is the donor’s transition dipole moment. It is convenient to determine the dipole moment of a real donor material from its absorption cross-section \cite{Novotny12principles}. For the sake of generality and to assess the performance of the optical platform, we normalized the dipole moment, therefore considering perfect donors and acceptors.
Within macroscopic QED, the resulting field at the acceptor’s position is $\hat{\mathbf{E}}(\mathbf{r}_A) = \frac{\omega^2}{\varepsilon_0 c^2} \, \tilde{\mathbf{G}}(\mathbf{r}_A, \mathbf{r}_D; \omega) \cdot \boldsymbol{\mu}_D \, \hat{\sigma}_D$.  
For the donor initially excited ($\langle \hat{\sigma}_D \rangle = 1$), this reduces to the classical dipole field:  
$\mathbf{E}(\mathbf{r}_A) = \frac{\omega^2}{\varepsilon_0 c^2} \, \tilde{\mathbf{G}}(\mathbf{r}_A, \mathbf{r}_D; \omega) \cdot \boldsymbol{\mu}_D$. The effective DDI potential energy $V_{\rm dd}$ arises from the exchange of a virtual photon during the joint transition  
$| e_D, g_A \rangle \;\longrightarrow\; | g_D, e_A \rangle$,  
where the donor de-excites and the acceptor is excited. Substituting $\hat{\mathcal{H}}_{\rm int} = -\hat{\boldsymbol{\mu}}_A \cdot \hat{\mathbf{E}}(\mathbf{r}_A)$ and evaluating the frequency integral yields \cite{Novotny12principles}:
\begin{equation}
    V_{\rm dd}(\mathbf{r}_A, \mathbf{r}_D; \omega) = \frac{\omega^2}{\varepsilon_0 c^2} \boldsymbol{\mu}_A \cdot \tilde{\mathbf{G}}(\mathbf{r}_A, \mathbf{r}_D; \omega) \cdot \boldsymbol{\mu}_D.
    \label{dyadic_GF_3}
\end{equation}
Here, $\boldsymbol{\mu}_D$ and $\boldsymbol{\mu}_A$ are the donor and acceptor dipole moments, and the real and imaginary parts of the DGF govern coherent and dissipative dynamics, respectively \cite{GonzalezTudela24}. Off-resonant terms enable nonlinearities and entangling gates, while near-resonant decay leads to super- and subradiant states that can guide or protect quantum information \cite{GonzalezTudela24}. The emitter’s decay rate is determined by $\mathrm{Im}\{\tilde{\mathbf{G}}\}$ \cite{Huidobro12}. In hyperbolic media, the DDI potential can also be written as  
$V_\text{dd} = \hbar \left( J_\text{dd} - i \frac{\gamma_\text{dd}}{2} \right)$,  
where $J_\text{dd}$ represents coherent energy exchange exchange through virtual photons and $\gamma_\text{dd}$ accounts for super- or subradiant effects \cite{Cortes17super}.

As a representative MIR hyperbolic material, we consider a 200-nm-thick $\alpha$-MoO\textsubscript{3} slab, which supports highly confined TM-polarized PhPs \cite{AlvarezPerez19analytical} (Fig.~\ref{fig2}a). Since s-SNOM measurements \cite{chen19modern} mainly probe $p$-polarized components, we analyze two $z$-polarized dipoles. We define $G_{zz}$ and $G_{zz}^0$ as the dyadic Green’s functions (DGFs) in the $\alpha$-MoO\textsubscript{3} slab and in vacuum, respectively. Following ref.~\cite{MartinSanchez21focusing}, we model the slab as a 2D conductivity layer (valid when $d \ll \lambda_{\rm PhP}$ \cite{AlvarezPerez19analytical}), which avoids calculating fields inside the slab. The two-point DGF reads:
\begin{widetext}
\begin{equation}
G_{zz}(\mathbf{r}_A, \mathbf{r}_D) = \frac{\frac{3}{\varepsilon^2} (\varepsilon_x^2 \Delta y^2 + \varepsilon_y^2 \Delta x^2) e^{-q_p k_0 H}}{\sqrt{\pi d} \ \varepsilon_x^2 \varepsilon_y^{5/4} (k_0 d)^2 (\varepsilon_x \Delta y^2 + \varepsilon_y \Delta x^2)^{5/4}}
e^{ i \left( \frac{2 \varepsilon \sqrt{ \varepsilon_x \Delta y^2 + \varepsilon_y \Delta x^2 }}{\varepsilon_x \sqrt{\varepsilon_y} d} - \frac{\pi}{4} \right) },
\label{eq:Gzz}
\end{equation}
\end{widetext}
where $k_0=\omega/c$ and $q_p=k_p/k_0$ is the normalized in-plane momentum \cite{MartinSanchez21focusing}:
\begin{equation}
q_p = \frac{-2\varepsilon \sqrt{\varepsilon_x^2 \Delta y^2 + \varepsilon_y^2 \Delta x^2}}{\varepsilon_x \sqrt{\varepsilon_y k_0 d} \sqrt{\varepsilon_x \Delta y^2 + \varepsilon_y \Delta x^2}}.
\label{eq:Gzz_wavevector}
\end{equation}

The DDI potential energy enhancement, $F_{\rm DDI}$, can then be calculated as
\begin{equation}
F_{\rm DDI}=\frac{|V_{\mathrm{dd}}(\textbf{r}_A,\textbf{r}_D; \omega)|}{|V_{\mathrm{dd}}^0(\textbf{r}_A,\textbf{r}_D; \omega)|}= \frac{|G_{zz}(\textbf{r}_A,\textbf{r}_D; \omega)|}{|G_{zz}^0(\textbf{r}_A,\textbf{r}_D; \omega)|},
\label{dyadic_GF_3}
\end{equation}
and the energy transfer rate scales as $\Gamma_{\rm ET}/\Gamma_{\rm ET}^0 \propto F_{\rm DDI}^2$ \cite{Huidobro12}.

\bibliography{achemso-demo}
\end{document}